\documentclass[11pt]{article}
\usepackage{a4wide}
\usepackage{amsmath,amssymb}
\usepackage{enumerate}
\usepackage{setspace}
\usepackage{footnpag,times,nicefrac}
\usepackage{graphicx}
\long\def\symbolfootnote[#1]#2{\begingroup%
\def\thefootnote{\fnsymbol{footnote}}\footnote[#1]{#2}\endgroup}
\def\di{\mathrm{d}}

\def\ci{\mathbb{C}}
\def\di{\mathrm{d}}
\def\su{SU(2)}
\def\slr{SL(2,\mathbb{R})}

\def\Id{{\rm 1\kern-.28em I}}

\begin{document}


\vskip 2.5cm

\centerline{\LARGE Local models of heterotic flux vacua: spacetime and worldsheet aspects}
\vskip 3mm
\centerline{\small
{\it Proceedings of the XVIth European Workshop on String Theory 2010, Madrid, 14-18 June 2010}}

\vskip 1.6cm
\centerline{\bf Luca Carlevaro$^{\spadesuit}$ and Dan Isra\"el$^{\diamond}$
\symbolfootnote[2]{Email: Luca.Carlevaro@cpht.polytechnique.fr,israel@iap.fr}}
\vskip 0.5cm

\vskip 0.3cm
\centerline{ \sl $^\spadesuit$ 
LAREMA, Universit\'e d'Angers,
2 Bd Lavoisier, 49045 Angers,} 
\centerline{Centre de Physique Th\'eorique, Ecole Polytechnique,
 91128 Palaiseau, France
}

\vskip 0.3cm
\centerline{ \sl $^\diamond$GRECO, Institut d'Astrophysique de Paris,
98bis Bd Arago, 75014 Paris, France\footnote{Unit\'e mixte de Recherche
7095, CNRS -- Universit\'e Pierre et Marie Curie}
}

\vskip 1.4cm

\centerline{\bf Abstract} \vskip 0.5cm 
We report on some recent progress in understanding heterotic flux compactifications, 
from  a worldsheet  perspective mainly. We consider local models consisting in torus fibration over warped Eguchi-Hanson space and 
non-K\"ahler resolved conifold geometries. We analyze the supergravity solutions and define a {\it double-scaling limit} of the 
resolved singularities, defined such that the geometry is smooth and weakly coupled. We show that, remarkably, 
the heterotic solutions admit solvable worldsheet CFT descriptions in this limit. This allows in particular to 
understand the important role of worldsheet non-perturbative effects.



\section{Introduction}
Heterotic compactifications with flux have attracted recently lot of interest, as a promising way of getting phenomenologically 
viable string vacua. Supersymmetric solutions have to solve the so-called 'Strominger system', known for a long time~\cite{Strominger:1986uh} 
(see also~\cite{Hull:1986kz,Becker:2002sx,Curio:2001ae,Louis:2001uy,LopesCardoso:2003af,Becker:2003yv,
Becker:2003sh,Becker:2005nb,Benmachiche:2008ma} later developments). However, having a non-trivial $\mathcal{H}$-flux results in the metric 
loosing K\"ahlerity and being conformally balanced instead of Calabi-Yau~\cite{Mich,Ivanov:2000ai,LopesCardoso:2002hd}.  An additional complication 
comes from anomaly cancellation, which requires satisfying the  generalized Bianchi identity, 
that uses the spin connection with torsion:
\begin{equation}\label{bianchi}
\di \mathcal{H} = \alpha' (\text{tr}\, \mathcal{R} (\Omega_{-}) \wedge\, \mathcal{R}(\Omega_{-})-\text{Tr}\, \mathcal{F}\wedge \mathcal{F})\, .
\end{equation}

A proof of the existence of a family of smooth solutions to the Bianchi identity has only appeared recently, for the
$T^2 \hookrightarrow \mathcal{M} \to K3$ supersymmetric vacua found in~\cite{Dasgupta:1999ss} using dualities. A proof was 
given in~\cite{Fu:2006vj} using the Hermitian connection, but can be made simpler using another choice of connection as 
explained in~\cite{Becker:2009df}. The fibration induces a warping of the  base and the appearance of torsion. 
Those backgrounds can be equipped with a gauge bundle that is the tensor product of an Hermitian-Yang-Mills bundle over the $K3$ base with 
an holomorphic line bundle on $\mathcal{M}$. With a non-compact Eguchi-Hanson base, this construction was made fully explicit~\cite{Fu:2008ga}.  
The warp factor was computed order by order in a $\alpha'/(g_s a)^2$ expansion, where $a$ stands for the resolution parameter.

As is shown in~\cite{Carlevaro:2008qf,Carlevaro:2009jx,Carlevaro:xxx}, another regime of interest is when the blow-up parameter $a$ is significantly smaller (in string units) than the total fivebrane charge. Then, one can define a sort of 'near-bolt' geometry, that describes the latter in the neighborhood of the resolved singularity. This region can  be decoupled from the asymptotic Ricci-flat region by defining a {\it double scaling limit}  which sends the asymptotic string coupling $g_s$ to zero, while keeping the ratio $g_s/ a$ fixed in string units.  In this limit the solutions are still weakly coupled and weakly curved, hence amenable to a supergravity description.

We have also found in~\cite{Carlevaro:2009jx} a new family of supersymmetric smooth torsional solutions, where 
the singularity of an orbifoldized conifold is smoothed out by a four-cycle, again supported by an Abelian gauge flux. The supersymmetry 
equations can be solved numerically in the asymptotically Ricci-flat case. Furthermore, in a double-scaling limit analogous to 
the previous example, one obtains an analytic solution, solving the Bianchi identity at leading order.

Heterotic torsional geometries, having only NSNS three-form and gauge fluxes, are expected to allow for a solvable worldsheet  description. By taking the double-scaling limit, as defined above, of torsional Eguchi-Hanson or resolved conifold geometries,  we have shown in~\cite{Carlevaro:2008qf,Carlevaro:2009jx} that the corresponding worldsheet non-linear sigma models admit  solvable worldsheet \textsc{cft} descriptions, belonging to a particular class of gauged \textsc{wzw} models.

The existence of a  worldsheet \textsc{cft} first implies that these backgrounds are exact heterotic string vacua to all orders in $\alpha'$, once included the worldsheet quantum corrections to the defining gauged \textsc{wzw} models; in particular, the corrected background will be a solution of the modified Bianchi identity. This proof of existence is particularly welcomed since the supergravity backgrounds  as they are are not generically an exact solution of the Bianchi identity, rather give an approximate solution of the latter in the limit where the fivebrane charge is large.

Having a full control over the \textsc{cft}  provides also information about worldsheet instanton corrections. These worldsheet  non-perturbative effects are captured by  Liouville-like interactions correcting the sigma-model action.  We analyze under which conditions these worldsheet operators are compatible with the whole construction (in particular with the orbifold  and~\textsc{gso} projections). This allows to understand several constraints  satisfied by consistent heterotic vacua from a worldsheet perspective, such as the tadpole condition, the evenness of the first Chern class, charge quantization and the corresponding moduli stabilisation.

\section{The supergravity solutions}
\label{secsugra}
Let us consider first the heterotic supergravity vacua corresponding to the local torsional supersymmetric compactifications. 
While the Eguchi-Hanson solutions were known, the conifold ones are new. 
\subsection{Fibrations over Eguchi-Hanson}
The $T^2 \hookrightarrow \mathcal{M} \to CY_2$ supersymmetric vacua have a metric of the form
\begin{equation}\label{metric2}
\di s^2_6  = \eta_{\mu\nu}\,\di x^\mu \di x^\nu +  \frac{\alpha' U_2}{T_2}\left|\di x^1 + T \di x^2+ (w^1 + w^2 T)
\mathcal{A} \right|^2 + 
e^{2(\Phi-\Phi_\infty)} \,\di s^2_\textsc{eh}\,. 
\end{equation}
where $U$ and $T$ are the K\"ahler and complex structure moduli of the torus and $w^{1,2}$ are the Kaluza-Klein charges of the fibration.  The warped non-compact base  is chosen to be Eguchi-Hanson space~\cite{Eguchi:1978xp}:
\begin{equation}\label{EHmetric}
\di s^2_\textsc{eh} =  \frac{ \di r^2}{1-\tfrac{a^4}{r^4}} +
\frac{r^2}{4} \left(
(\sigma^\textsc{l}_1)^2 + (\sigma^\textsc{l}_2)^2 +
\Big(1-\frac{a^4}{r^4}\Big) (\sigma^\textsc{l}_3 )^2 \right)\,,
\end{equation}
here written in terms of the left-invariant $SU(2)$ one-forms. The existence of a unique anti-selfdual $(1,1)$-form on EH dictates the 
choice of the connexion one-form on the base
\begin{equation}
\mathcal{A} = \frac{a^2}{2r^2} \sigma^\textsc{l}_3 \, ,
\end{equation}
preserving $\mathcal{N}_{\textsc{st}}=2$ supersymmetry. One can add a supersymmetric Abelian gauge bundle of the form $\mathcal{F} =  \vec{\ell} \cdot \vec{\mathfrak{T}}\, \di \mathcal{A}$,  where $\mathfrak{T}^{i}$ are the Cartan generators of $\mathfrak{so}(32)$ and the vector $\vec{\ell}$ encodes the magnetic charges.

The heterotic Bianchi identity cannot be solved easily. One can consider an approximate solution of~(\ref{bianchi}) by 
taking a {\it large charge limit}~\cite{Carlevaro:2008qf}.  This is possible $e.g$ for $\vec{\ell}^2\gg 1 $. We have then 
\begin{equation}\label{bianchilow}
\di \mathcal{H} -\alpha ' \,\text{Tr}
    \mathcal{F} \wedge \mathcal{F} + \mathcal{O}\big(1) = 0\, ,
\end{equation}
and the charges should satisfy the tadpole condition
\begin{equation}\label{tad}
\mathcal{Q}_5 - \frac{U_2}{2T_2} |w^1 + w^2 T|^2 -  \vec{\ell}^{\,2}= 0 \,.
\end{equation}

\subsection{Resolved conifold}
As a second example we consider a warped conifold geometry~\cite{Candelas:1989js}. The singularity is resolved by a K\"ahler deformation corresponding to blowing up a  $\ci P^1\times \ci P^1$ four-cycle. This is topologically possible only for a $\mathbb{Z}_2$ orbifold of the conifold.  The geometry  is conformal to a six-dimensional smoothed cone over a 
non-Einstein $T^{1,1}$ space~:
\begin{subequations} \label{sol-ansatz}
\begin{align}
\di s^2 & = \eta_{\mu\nu}\di x^{\mu}\di x^{\nu} +\tfrac{3}{2}\,e^{\Phi-\Phi_\infty} \,\biggl[
\frac{\di r^2}{f^2(r)} + 
 \frac{r^2}{6}\big(\di\theta_1^2+\sin^2\theta_1\,\di\phi_1^2 + \di\theta_2^2+\sin^2\theta_2\,\di\phi_2^2\big) \notag\\
 & \hspace{2cm} +\,\frac{r^2}{9} f(r)^2 \big(\di \psi + \cos \theta_1 \,\di \phi_1 + \cos \theta_2 \,\di \phi_2 \big)^2  \biggr]\,,\\
\mathcal{H} &  = \frac{\alpha'k}{6}\,g_1(r)^2\,\big( \Omega_1+\Omega_2\big)\wedge  \tilde{\omega}\,,
\end{align}
\end{subequations}
with the volume forms of the two $S^2$s and the connection one-form $\tilde{\omega}$ defined by
\begin{subequations} 
\begin{align}
\Omega_i&=\sin \theta_i \,\di \theta_i\wedge \di \phi_i\,,\quad \text{for }i=1,2\,,\qquad
\tilde{\omega} = \di \psi + \cos \theta_1 \,\di \phi_1 + \cos \theta_2 \,\di \phi_2\,.
\end{align}
\end{subequations}
One can consider again an Abelian gauge bundle, supported both on  the four-cycle $\ci P^1\times \ci P^1$  at the bolt and on a 
non-compact cycle:
\begin{equation}\label{gauge-ansatz}
\mathcal{A}=\tfrac{1}{4}\Big(\left(\cos\theta_1\,\di\phi_1 - \cos \theta_2\,\di \phi_2 \right)\vec{p} + g_2(r)\,\tilde\omega\,\vec{q}\Big) \cdot \vec{\mathfrak{T}}\,.
\end{equation}
characterized this time by two charge vectors $\vec{p}$ and $\vec{q}$, that should be orthogonal. 

To determine the radial profile of the three-form $\mathcal{H}$, $i.e.$ the function $g_2(r)$ in the ansatz~(\ref{sol-ansatz}), we  solve the Bianchi identity in the large charges limit $\vec{p}^{\, 2} \gg 1$. We obtain:
\begin{equation}\label{g3}
g_1^2(r)=\tfrac34\big[1-g_2^2(r)\big]= 
\tfrac34\Big[1-\left(\frac{a}{r}\right)^8\Big]
\end{equation}
with the relation $\vec{p}^{\, 2} =  \,\vec{q}^{\, 2} = k$. 

Having analytical expressions for the functions $g_1$ and $g_2$, we can consider solving the first order supersymmetry 
equations for the remaining functions $f$ and $\Phi$.  If we ask the conformal factor $e^{\Phi}$ to be asymptotically constant, 
they can only be solved numerically.

\subsection{Double-scaling limit}
An important quantity is the value of the conformal factor evaluated at the bolt $r=a$; when it is large, the resolved 
cycle sits deep in the region of strong warping. One can then consider the flux background corresponding to the 'near-bolt' geometry.  

Let us start with the Eguchi-Hanson case. In order to isolate the region near the two-cycle of the resolved $A_1$ singularity, 
where the interesting physics takes place, one  can define a certain 
{\it double-scaling limit} of the the type \textsc{ii} and heterotic solutions as follows:
\begin{equation}\label{DSL}
g_s \to 0 \quad , \qquad \lambda= \frac{g_s \sqrt{\alpha'}}{a} \quad \text{fixed}\,,
\end{equation}
with $g_s = e^{\Phi_\infty}$. The effective  string coupling constant, set by the double-scaling parameter $\lambda$, stays finite and can chosen within the perturbative regime. In terms of a new radial coordinate defined by $\cosh \rho = (r/a)^2$, 
which is held fix in this limit, on obtains the metric, independently of $a$, as
\begin{equation}\label{metric3}
\di s^2 = \di x^\mu \di x_\mu + \frac{\alpha' U_2}{T_2} \left| \di y^1 + 
T \di y^2 +  \frac{\mathfrak{q}\,\sigma_3^{\textsc{l}}}{2\cosh\rho} \right|^2 + 
\frac{\alpha' Q_5}{2}\Big[ {\rm d}\rho^2 + (\sigma^\textsc{l}_1)^2 +
(\sigma^\textsc{l}_2)^2 + \tanh^2 \rho \, (\sigma^\textsc{l}_3 )^2\Big]\,.
\end{equation}

In the conifold case,  the double-scaling limit is slightly different, as now 
$\mu=\frac{g_s \alpha'}{a^2}$ is held fixed. It again isolates the dynamics  near the four-cycle of the resolved singularity, keeping 
the transverse space to be conformal to the non-singular resolved orbifoldized conifold. The supersymmetry equations can in this case 
be solved analytically:
\begin{equation}\label{Hfg}
e^{\Phi} =\mu  k \left(\frac{a}{r}\right)^2\,,\qquad f^2(r)=g_1^2(r)=\tfrac{3}{4}\Big[1-\left(\frac{a}{r}\right)^8\Big]\, .
\end{equation}

\section{The worldsheet description}
A remarkable feature  of the $T^2 \hookrightarrow \mathcal{M} \to EH$ and resolved 
orbifoldized conifold solutions is that they admit a solvable worldsheet CFT description in the double-scaling limit. 
We start by discussing the former. 
\subsection{Blow-down limit}
It is easier to start with the blow-down limit of the background, such that the Abelian instanton becomes point-like. 
In this limit the solution  boils down to the near-horizon five-brane solution (the so-called \textsc{chs} background~\cite{Callan:1991dj}), transverse to a 
$\mathbb{C}^2/\mathbb{Z}_2$ orbifold and compactified on a two-torus:
\begin{equation}
\di s^2 = \di x_\mu \di x^\mu +\tfrac{U_2}{T_2}\left|\di x_1 + T \di x_2 \right|^2 + \alpha' k \left[ 
\di \rho^2 + \di s^2 (S^3/\mathbb{Z}_2) \right]\, ,
\end{equation}
that features a linear dilaton  and a non-zero NSNS three-form flux.  
We consider the case of $Spin(32)/\mathbb{Z}_2$ heterotic strings.  The worldsheet CFT is 
\begin{equation}
\mathbb{R}^{2} \times T^2 \times \mathbb{R}_Q  \times SU(2)_{k}/\mathbb{Z}_2 \times 
\left. SO(32)_1 \right|_\textsc{r}
\label{blowdowncft}
\end{equation}
The background charge of the linear dilaton, $Q=\sqrt{2/k}$, is related to the level $k$ of the supersymmetric $SU(2)$ WZW model.  
$k$ has to be even in order to allow for a $\mathbb{Z}_2$ orbifold of this WZW model.  

\subsection{Blown-up theory and Liouville potentials}
In order to get the resolved background it is easier to use a dual formulation 
(generalizing~\cite{fzz}) of the deformation. We add to the conformal field theory~(\ref{blowdowncft}) the following generalized Liouville interaction~\cite{Carlevaro:xxx}
\begin{equation}
\label{Liouvillepotgenericgeneric}
\delta S = \mu \int \di^2 z \,  \left\{ G_{-1/2}, \, e^{-\sqrt{k}\left(\varrho (z,\bar z) + i Y_L (z) \right)}\right\}   
e^{i \tilde{p}_a \tilde{X}^a (\bar z) + i \vec{\ell} \cdot \vec{Z}_R (\bar z)} + c.c.
\end{equation}
As discussed below, the target-space of the corresponding non-linear sigma model correspond indeeds to~~(\ref{metric3}). One can consider either an operator belonging to the right NS sector, with $\vec{\ell} \in \mathbb{Z}^{16}$, 
or to the right Ramond sector, with $\vec{\ell} \in (\mathbb{Z}+\tfrac{1}{2})^{16}$. The former case corresponds to bundles with 
vector structure, and the latter to bundles without vector structure. 

In order to get this marginal deformation from the physical spectrum of the blow-down theory, the  lattice of the two-torus 
should contain an element with
\begin{subequations}
\label{chiralpot}
\begin{align}
p_{a} &= 0 \quad , \qquad a = 1,2 \label{condrat1}\\
G_{ab}\tilde{p}^a \, \tilde{p}^b +2\vec{\ell}^2 &=k+2 \label{condrat2}
\end{align}
\end{subequations}
The first condition~(\ref{condrat1}) implies that the underlying $T^2$ conformal field theory is rational 
(as it is characterized then by a having chiral algebra). For this the torus moduli $T$ and $U$ need to be valued in the same {\it maginary quadratic number field}~\cite{Gukov:2002nw}. The second condition~(\ref{condrat2}) is the worldsheet version of the integrated space-time Bianchi identity, and  is exact to all orders in $\alpha'$. 

In order to satisfy the right GSO projection, the right fermion number of the operator in~(\ref{Liouvillepotgenericgeneric}) should be even ($i.e.$ that $\sum_i \ell_i \equiv 0 \mod 2$). This constraint corresponds exactly to the condition that the 
first Chern class of the gauge bundle belongs to the even integral second cohomology group,  found in~\cite{Witten:1985mj,Freed:1986zx}.

Furthermore, the operator~(\ref{Liouvillepotgenericgeneric}) belongs necessarily to the twisted sector of the 
$SU(2)/\mathbb{Z}_2$ orbifold. Hence, as in supergravity, one should have the identification $\psi \sim \psi +2 \pi$.

\subsection{The conifold case}
Let us consider briefly the conifold case. In the blow-down limit, one gets a linear dilaton, together with a non-Einstein $T^{1,1}$ space. The latter is described by an asymmetric gauged WZW model $U(1)\backslash (SU(2)_k\times SU(2)_k)$. This gauging, acting 
only by left multiplication, is anomalous. The anomaly is cancelled by coupling minimally the 32 right worldsheet fermions 
of $SO(32)_1$. This corresponds in spacetime to the gauge bundle specified by the vector of magnetic charges $\vec{p}$ in eq.~(\ref{gauge-ansatz}).  One considers a $\mathbb{Z}_2$ supersymmetric orbifold of the conifold, corresponding to  the half-period shift $\psi \sim \psi +2 \pi$, 
as for Eguchi-Hanson. 

The CFT for the resolved conifold solution~(\ref{sol-ansatz}), with a generic Abelian gauge bundle over the resolved conifold, is obtained from the blow-down theory $\mathbb{R}_Q \times T^{1,1} \times SO(32)_1 |_\textsc{r}$ by the marginal deformation
\begin{equation}
\label{Liouvintgen}
\delta S = \mu_\textsc{l} \int \di^2 z (\psi^\rho +i\psi^\textsc{x})e^{-\frac{\sqrt{\vec{q}^{\,2}-4}}{2} (\rho + i X_\textsc{l})} 
\, e^{\frac{i}{2}\vec{q} \cdot \vec{Y}_\textsc{r}}
+ c.c. \,.
\end{equation}
Again we require this operator to be part of the physical spectrum of the blow-down theory, 
taking into account the \textsc{gso} and orbifold projections. 

We consider bundles with $\vec{p}^{\, 2}\equiv 2 \mod 4$, for which no action in the $Spin(32)/\mathbb{Z}_2$ lattice is 
requested by orbifold invariance. We find that the operator in~(\ref{Liouvintgen}) is GSO-invariant provided that 
\begin{equation}
\frac{1}{2}\sum_{i=1}^{16} (q_i \pm  p_i) \equiv 0 \mod 2 \,.
\label{fwcondcft}
\end{equation}
This condition  is again the same as $c_1 (\mathcal{F}) \in H^2(\mathcal{M}_6,2\mathbb{Z})$. This remarkable 
relation between  topological properties of the gauge bundle and the \textsc{gso} parity of worldsheet instanton corrections may 
originate from modular invariance, that relates the existence of spinorial representations of the gauge group to the projection with the right-moving worldsheet fermion number.

\subsection{The gauged WZW model approach}
The algebraic construction of the worldsheet \textsc{cft} outlined above gives, in both cases, non-linear sigma-models 
whose background fields are identical, at leading order in $\alpha'$, to the solutions given in section~\ref{secsugra}. 
It is shown by using a complementary description of these worldsheet theories in terms of gauged WZW models. 

For instance, theCFT with warped deformed orbifoldized conifold as  target space is given by the asymmetric $\mathcal{N}_{\textsc{ws}}=(1,0)$  super-coset: 
\begin{equation}
\frac{\slr_{k/2} \times\, \left(\text{\small \raisebox{-1mm}{$U(1)_\textsc{l}$}\! $\backslash$ \!\!\raisebox{1mm}{$SU(2)_k \times SU(2)_k$}}\right)\times SO(32)_1 |_\textsc{r}}{U(1)_\textsc{l} \times U(1)_\textsc{r}} \,,
\label{cosetdef}
\end{equation}
which combines a left gauging of $SU(2) \times SU(2)$ with a pair of chiral gaugings which also involve an $\slr$ \textsc{wzw} model. The whole coset is made anomaly-free by coupling the $SO(32)_1$ right-moving gauge sector. The total action for the gauged \textsc{wzw} model defined above is given schematically as follows:
\begin{equation}\label{Stot}
S_{\textsc{wzw}}(A,{\bf B}) = S_{\slr_{k/2+2}} + S_{\su_{k-2},\, 1} +  S_{\su_{k-2},\, 2} + S_\text{gauge}(A,{\bf B}) + S_\text{Fer} (A,{\bf B})\, ,
\end{equation}
where the first three factors correspond to bosonic \textsc{wzw} actions, the fourth one to the bosonic terms involving the gauge fields and the last one to the action of the minimally coupled fermions realizing the $\mathfrak{so}(32)_1$ algebra. 

Finding the background fields, at leading order in $\alpha'$, amount then to integrate out classically 
the worldsheet gauge fields $A$ and ${\bf B}$. For  heterotic cosets this is in general more tricky  because of the worldsheet anomalies generated by the various pieces of the model~\cite{Johnson:1994jw,Johnson:2004zq}. When the dust settles
we eventually find a heterotic sigma-model corresponding, as promised, to the torsional heterotic solutions of interest.

\section{Conclusions}
We have analyzed local models of smooth torsional heterotic compactifications preserving 4d space-time supersymmetry. 
From the supergravity viewpoint we have found a class of conifold solutions, that are the first explicit examples of genuinely 
$SU(3)$ structure heterotic solutions. We have argued that one can define a double-scaling limit that focuses on the 
resolved singularities of these manifolds

In this limit, we have shown that the associated worldsheet theory becomes a fully solvable conformal field theory. We have found 
an interesting interplay between worldsheet non-perturbative effects, corresponding to Liouville-like interactions, and constraints 
on heterotic vacua usually considered in supergravity: charge quantization and associated moduli stabilization, evenness of the 
first Chern class, or avoidance of conical singularity at a bolt. 

Among the possible followups of these work, it would be extremely interesting to compute the perturbative corrections to these solutions, using the methods of~\cite{Tseytlin:1992ri,Bars:1993zf}, which would give a corrected background {\it to all orders} in $\alpha'$. In particular, it would give exact solutions to the modified Bianchi identity.

It is highly desirable to consider not only local models, but genuine compactifications with torsion. In~\cite{Adams:2006kb} linear 
sigma-models of $T^2 \hookrightarrow \mathcal{M} \to K3$ where constructed. Using the methods described here, one should be able to 
obtain explicit conformal field theories corresponding to their infrared fixed points. This would give solid grounds for extracting 
low-energy four-dimensional physics out of heterotic flux compactifications.


\bibliography{bibbundle}

\end{document}